\documentclass[aps,twocolumn,amsmath,amssymb,superscriptaddress,prl,letter]{revtex4}
\usepackage{graphicx}

\begin{document} 
\title{
Fermionic Shadow Wavefunction Variational calculations of the vacancy formation energy in $^3$He.
} 

\author{L. Dandrea}
\affiliation{Dipartimento di Fisica, University of Trento,
via Sommarive 14, I-38050 Povo, Trento, Italy}
\affiliation{INFN, Gruppo Collegato di Trento, Trento, Italy}
\author{F. Pederiva}
\affiliation{Dipartimento di Fisica, University of Trento,
via Sommarive 14, I-38050 Povo, Trento, Italy}
\affiliation{INFN, Gruppo Collegato di Trento, Trento, Italy}
\author{S. Gandolfi} 
\affiliation{S.I.S.S.A., International School of Advanced Studies,
via Beirut 2/4, 34014 Trieste, Italy}
\affiliation{INFN, Sezione di Trieste, Trieste, Italy}
\author{M. H. Kalos}
\affiliation{Lawrence Livermore National Laboratory Livermore, CA 94550 USA}

\begin{abstract}
We present a novel technique well suited to study the 
ground state of inhomogeneous fermionic matter 
in a wide range of different systems. The system is 
described using a Fermionic Shadow wavefunction (FSWF)
and the energy is computed by means of the Variational 
Monte Carlo technique. The general form of FSWF is 
useful to describe many--body systems with the coexistence 
of different phases as well in the presence of 
defects or impurities, but it requires overcoming a significant 
sign problem.  As an application, we studied the energy 
to activate vacancies in solid $^3$He.
\end{abstract}

\maketitle

The microscopic theoretical description of inhomogeneous Fermionic systems
is a long--standing challenge. Among such systems we can include
defective quantum crystals (e.g. $^3$He crystals or electron
Wigner crystals in presence of vacancies or defects) or where
an ordered and a disordered phase coexist like, for instance, fluid and crystal.
The main difficulty consists in the fact that one has to deal with a wavefunction
that combines the antisymmetry required by the Pauli principle
and the inhomogeneity of the system itself. 

While mean field methods are very efficient in dealing with homogeneous
phases (e.g. an extensive and perfect solid), the phase coexistence or
the description of local defects presents difficulties. The main reason is
the locality of the inhomogeneity. Therefore, 
an explicit description of the wavefunction seems a much better approach.

The rigorous microscopic evaluation of the vacancy formation energy
in $^{3}$He is one of the problems that most suffers from the limitations
of standard theoretical tools.

The problem was successfully solved for many-Boson systems several years ago,
by means of the so-called Shadow Wave Functions (SWF)\cite{vitiello88,Reatto88,vitiello90}, a class of wavefunctions
based on the introduction of auxiliary degrees of freedom, which was 
successfully applied to a variety of inhomogeneous phases of $^{4}$He\cite{pederiva94,Galli08} and
p-H$_{2}$\cite{operetto07}:
\begin{equation}
\label{eq:SWF}
\psi_{SWF}(\textbf{R})=\phi_p(\textbf{R})\int \Xi(\textbf{R}, \textbf{S})\phi_s( \textbf{S}) d\textbf{S} \,,
\end{equation}
where $\textbf R=\{\bf{r}_1,...,\bf{r}_N\}$ are the coordinates of the $N$ constituents
of the system, and
$\textbf S=\{\bf{s}_1,...,\bf{s}_N\}$ are auxiliary degrees
of freedom called ``shadows''. $\phi_p$ and $\phi_s$ are 
two--body correlation factors (of the so-called Jastrow form) for
particles and shadows, respectively, and $\Xi(\textbf R,\textbf S)$ is the kernel 
describing the correlations between particles and shadows. 
Here, as in most applications, we take the kernel to be a Gaussian:
\begin{equation}
\label{eq:GKER}
\Xi(\textbf{R}, \textbf{S})=\exp(-c(\textbf R -\textbf S)^2) .
\end{equation}

The main properties of the SWF are i) the fact that it introduces correlations
to all orders via the integration over the auxiliary degrees of freedom
and ii) the fact that despite its manifest translational invariance, it can
describe phases in which the translational symmetry is broken (solids,
interfaces, defects).
However, the extension to many-Fermion systems is hard.
A straightforward extension (which we term ASWF) based on the antisymmetrization of the particle 
degrees of freedom was proposed several years ago for the study
of homogeneous $^{3}$He\cite{pederiva96}, and later applied to the homogeneous electron 
gas\cite{dandrea06}:
\begin{eqnarray}
\psi_{ASWF}(\textbf R) &=&  
\prod_{l=\uparrow,\downarrow}D_l[\phi_k(\textbf r_i)] \phi_p(\textbf R)
\nonumber \\
&\times& \int \exp(-c(\textbf R -\textbf S)^2) \phi_s(\textbf S) d\textbf S \,,
\end{eqnarray}
where $D_l[\phi_k(\textbf r_i)]$ is a Slater determinant of particle orbitals.

This form has an evident drawback. Once the orbitals are specified, as
{\it{e.g.}} plane waves satisfying the Born-von Karman
conditions for an extensive system, the nodal structure remains unchanged
even if the auxiliary degrees of freedom provide the correlations necessary to
break the symmetry so as to correctly describe a set of localized particles.

The only viable solution to the problem is to introduce an antisymmetric wavefunction
in which crystallization can be described without explicit symmetry breaking, but
that at the same time can develop a correct nodal structure according to the phase
described.
This goal can be achieved by writing a Shadow Wave Function in which the
antisymmetry is imposed on the auxiliary degrees of freedom, therefore
maintaining explicit correlations among the particles only in the symmetric
part of the function. The so-called Fermion-Shadow Wave Function (FSWF) 
assumes the following form:

\begin{eqnarray} \label{eq:xi_fswf}
\psi_{FSWF}(\textbf R) &=& 
\phi_p(\textbf R)
 \int \exp(-c(\textbf R -\textbf S)^2) \nonumber \\
&& \times \prod_{l=\uparrow,\downarrow}D_l[\phi_k(\textbf s_i)]
\phi_s(\textbf S) d\textbf S \,.
\end{eqnarray}
It is possible to prove that $\psi_{FSWF}$ is antisymmetric under the exchange
of two particles of like spin \cite{Rea95}. The main difference between ASWF and
FSWF comes from the fact that the latter develops a nodal structure for the 
particle degrees of freedom that depends on the integration over the shadow
degrees of freedom, and includes effects of correlations to all orders. In particular
it is easily proved that when particles and shadows are strongly localized by the effect of 
the two--body correlations among the shadows themselves, the wavefunction
is closely approximated by a determinant of Gaussians connecting each particle to each shadow,
which is obviously closer to the structure of the wavefunction expected for a 
quantum crystal.

There is a very high technical price to pay in order to exploit FSWF in computations.
In a Variational Monte Carlo (VMC) calculation, the absolute square of the 
wavefunction is used as a probability density, $P(\textbf R)$, and the local energy is averaged 
over the sampled configurations.  Thus
\begin{equation}
\label{eq:ene}
E=\frac{\int d\textbf R\psi^*(\textbf R)\psi(\textbf R) E_L}{\int d\textbf R\psi^*(\textbf R)\psi(\textbf R)}
= \frac{\int d\textbf R P(\textbf R) E_L}{\int d\textbf R P(\textbf R)} \,,
\end{equation}
where $E_L=\psi^{-1}H\psi$ is the local energy of the system.
The integral is evaluated by generating configurations according to 
$P=|\psi|^2$ that are sampled 
using the Metropolis algorithm. 

In using SWF, one constructs $\psi^*(\textbf R)\psi(\textbf R)$ by integrating over
two sets of shadow variables, $\textbf S$ and $\textbf S'$.  Define the
integrand of SWF as
\begin{eqnarray}
\label{eq:DOI}
&& Z(\textbf R,\textbf S,\textbf S') =  
\phi^2_p(\textbf R)
\nonumber \\
&&\times \exp(-c(\textbf R -\textbf S)^2-c(\textbf R -\textbf S')^2) 
\phi_s(\textbf S) \phi_s(\textbf S') \,
\end{eqnarray}
and choose some probability density function, $\tilde P(\textbf R,\textbf
S,\textbf S')$. Sampling in the usual way,
a generic operator can be computed as
\begin{equation}
\label{eq:fop}
\langle O\rangle=\frac{\int d\textbf R d\textbf S d\textbf S' \tilde P(\textbf R,\textbf S,\textbf S') w(\textbf R,\textbf S,\textbf S') O(\textbf R)}
{\int d\textbf R d\textbf S d\textbf S' \tilde P(\textbf R,\textbf S,\textbf S')
w(\textbf R,\textbf S,\textbf S') } \,,
\end{equation}
where
\begin{equation}
\label{eq:dow}
w(\textbf R,\textbf S,\textbf S') = \frac{Z(\textbf R,\textbf S,\textbf S')}{\tilde P(\textbf R,\textbf S,\textbf S') }  .
\end{equation}
A reasonable choice for ordinary SWF is the integrand itself:
\begin{equation}
\label{eq:PSS}
P(\textbf R,\textbf S,\textbf S') =  Z(\textbf R,\textbf S,\textbf S') ,
\end{equation}
with $w=1$ and $\tilde P=P = Z$.
Similarly in $\psi_{ASWF}$ the integrand is positive, and can be used for 
$P(\textbf R,\textbf S,\textbf S')$.

When using  $\psi_{FSWF}$,  the integrand 
is not positive definite, and sampling it is not possible. It is always
possible, however, to sample a suitable probability distribution and
compute a weighted average. The most straightforward choice
in this case is the absolute 
value of the integrand. Let
\begin{eqnarray}
&&Q(\textbf R,\textbf S,\textbf S') =
\phi^2_p(\textbf R) \exp(-c(\textbf R-\textbf S)^2-c(\textbf R -\textbf S')^2) 
\nonumber \\
&&\times \prod_{l=\uparrow,\downarrow}D_l[\phi_k(\textbf s_i)]
\prod_{l'=\uparrow,\downarrow}D_l[\phi_k(\textbf s'_i)]
\phi_s(\textbf S) \phi_s(\textbf S') \,, 
\\
&&\tilde P(\textbf R,\textbf S,\textbf S') = |Q(\textbf R,\textbf S,\textbf S')| \,, 
\\
&&w(\textbf R,\textbf S,\textbf S') = \frac{Q(\textbf R,\textbf S,\textbf S')}{|Q(\textbf R,\textbf S,\textbf S')|} \equiv \pm 1 \,.
\end{eqnarray}
It should be noted that the normalization integral, while containing positive
and negative terms,
is always positive by construction. 
However, the speed of convergence of the integral strongly depends on the fluctuations
in sign of the integrand. In particular, the intrinsic variance of the integrand
might become excessively large, and prevent the computation of an average with 
acceptable statistical errors.
This is particularly true for disordered systems, such as liquids or disordered solids, where the
wavefunction has strong variations in space.

A simple reorganization of the calculation produces
a dramatic improvement in the Monte Carlo efficiency.
In applying the Metropolis method to shadow wavefunction, including
the ASWF variant, the usual procedure is to sample new values of 
$\textbf R$, $\textbf S$, and $\textbf S'$ in 
turn.  It is always true that the integrals in Eq. (\ref{eq:fop}) over $\textbf S$ and $\textbf S'$
for fixed $\textbf R$ are positive.  This suggests that a change in the
order of summation might be useful
for the FSWF class of functions, especially with disorder,  
by propagating
the shadows $\textbf{S}$ and $\textbf{S}'$ for $M$ steps (with $M$ big 
enough)  for fixed $ \textbf R$. 
That is, we expect that the sum of the $\pm1$ weight 
of $M$ steps,
\begin{equation}
\left[\sum w_i\right]_{S}\times\left[\sum w_i\right]_{S'}=W_S W_{S'} \,,
\end{equation}
will be usually positive.  In fact, increasing $M$
in the more difficult cases where exchanges 
of sign often appear, 
gives weights $W_S W_{S'}$ usually positive and significantly 
different from zero. 
The algorithm becomes: i) sample a configuration $\textbf R$ of 
particles, ii) sample $M$ 
configurations of $\textbf S$, iii) sample $M$ configurations of 
$\textbf S'$, iv) combine all 
the weight factors and accumulate the local energy for the 
average and variance, 
iterate from i) to iv) until the convergence is reached 
and the variance is low as desired.
For a calculation of the crystalline phase with no empty 
sites we typically sampled  configurations
of particles, and using $M=1$ or $M=100$ does not significantly 
change the result. If a vacancy
is present in the system and the sign frequently changes, 
we usually sampled $5\times 10^{6}$ configurations
and for each one $M=1.5\times 10^3$.

As an illustration of the capabilities of FSWF, we studied the ground 
state of  solid $^3$He in the 
range of molar volumes between 20 and 24 cc/mol.  A comparison
with previous variational estimates based
on standard antisymmetric wavefunctions is given below.
As a next step, we studied the vacancy 
formation energy by computing the energy
in the presence of an empty site. 
As already pointed out, standard imaginary time projection 
calculations require that a wavefunction imposing crystallization be used,
preventing one from taking properly into account all the phenomenology
related to lattice relaxation and vacancy mobility.
As model He--He interaction we chose the
Hartree-Fock dispersion HFDHE2 potential by Aziz \emph{et al.} \cite{Azi70}, which
gives an overall description of the equation of state in good agreement with
experiments, though it does not introduce explicit three-body terms.
In the Jastrow functions $\phi_s$ the pseudopotential $u_s(r_{ij})$ was taken as the rescaled
particle-pair potential, $v(r_{ij})$ namely
$u_s(r_{ij})=\alpha v(\beta r_{ij})$ ($\alpha$ and $\beta$ are additional 
variational parameters), while in the $\phi_p$ we
used a McMillan form\cite{pederiva96} combined with a summation over 
a basis as in Ref. \cite{moroni95}.
All the variational parameters entering in the wavefunction were 
optimized at each density
using an energy--variance minimization technique due to C.J. Umrigar and M.P Nightingale 
applied to the system with no defects. 

\begin{table}[h]
\begin{center}
\begin{tabular}{c|cc|cc}
\hline
$\rho\sigma^3$ & E/N(54)   & T/N(54)   & E/N(53)   & T/N(53) \\
\hline                                                      
0.419          & 0.422(1)  & 23.947(1) & 0.69(1)  & 23.71(3) \\
0.427          & 0.548(2)  & 24.608(1) & 0.78(8)  & 24.6(2)  \\
0.438          & 0.955(1)  & 26.005(2) & 1.26(1)  & 25.76(3) \\
0.457          & 1.556(1)  & 27.986(2) & 1.844(8) & 28.05(2) \\
0.479          & 2.455(1)  & 30.482(2) & 2.801(7) & 30.60(2) \\
0.503          & 3.481(1)  & 32.487(2) & 4.127(7) & 32.35(2) \\
\hline
\end{tabular}
\end{center}
\caption{Total and kinetic energy per particle at different densities in the solid b.c.c. phase for the
crystal with no defects (54 atoms) and with the presence of an empty site (53 atoms).
All the energies are expressed in $K$.}
\label{tab:ene}
\end{table}

\begin{figure}[ht]
\vspace{0.2cm}
\begin{center}
\includegraphics[width=7cm]{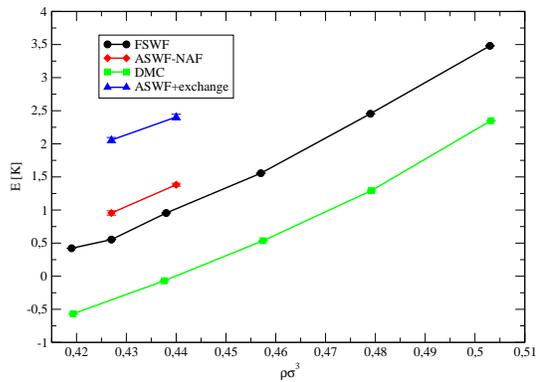}
\caption{(color online)
The FSWF energy per particle (black circles) as a function of the density. The result is compared with
the two results provided by ASWF of Ref. \cite{pederiva96} (blue triangles and red diamonds) and
with the DMC results of Ref. \cite{moroni00}.}
\label{fig:eos54}
\end{center}
\end{figure}

We report the energy of 54 atoms in table \ref{tab:ene}.
The energy per particle is also displayed in Fig. \ref{fig:eos54} 
where we compared our results (circles) with
those found in Ref. \cite{pederiva96} computed using ASWF
starting from a normal antiferromagnetic 
order NAF (diamonds), and including exchanges (triangles), and with the more accurate 
Diffusion Monte Carlo (DMC) results of Ref. \cite{moroni00} (squares).
As can be seen FSWF provides the lowest of the variational 
estimates of the energy. The DMC energies are 
lower at each density by a constant value of about 1K.

The vacancy formation energy at constant pressure for a system 
with $N$ particles at a 
fixed density $\rho$ can be computed as\cite{guyer71,pederiva97,operetto07}
\begin{equation}
\label{eq:enevac}
\Delta E_v=E(N-1,N_l=N)-\frac{N-1}{N}E(N,N_l=N) \,,
\end{equation}
where the number of lattice sites $N_l$ is conserved and 
the density of the two systems is the same.
The vacancy formation energy includes contributions from lattice relaxation
and tunneling that cannot be accounted for by a wavefunction
with an underlying lattice structure (such as a Jastrow-Nosanow
wavefunction). 
The computation of the energy for the system with $N-1$ 
particles is performed by removing one particle
and one shadow from the trial wavefunction. It has 
to be noted that dropping one shadow from the 
shadow determinant means having a hole state in one of 
the determinants of Eq. \ref{eq:xi_fswf}.
In the case of an open--shell configuration where one or more 
single particle states are not filled
one needs to perform the calculation using twist average boundary 
conditions \cite{Lin01,gandolfi09}, so that
the total wavefunction has zero total momentum.
Using FSWF, we conjecture that it is not important which particular 
shadow orbital is omitted, because the total momentum of the 
system is always conserved.
 
We tested this by repeating the same calculation
by removing different single--shadow states from the kernel. In 
particular the energy of the system 
where orbitals with different vector number $\textbf n$ were 
removed is the same within statistical error.
We stress that, by contrast, in using a normal many--body wavefunction 
with no shadows or using ASWF 
the energy would depend upon the unfilled single--particle 
orbital.  We assume for now that the system containing
one empty site is well described by the same wavefunction 
of the system with the complete crystal. The 
modified structure of the system with a vacancy 
is modeled by the shadow extra variables. 
Therefore for 53 atoms we used the same parametrization of 
$\phi_p$, $\phi_s$ and the coupling constant $c$ entering in
Gaussians of the system with 54 atoms. 

\begin{figure}[ht]
\vspace{0.2cm}
\begin{center}
\includegraphics[width=7cm]{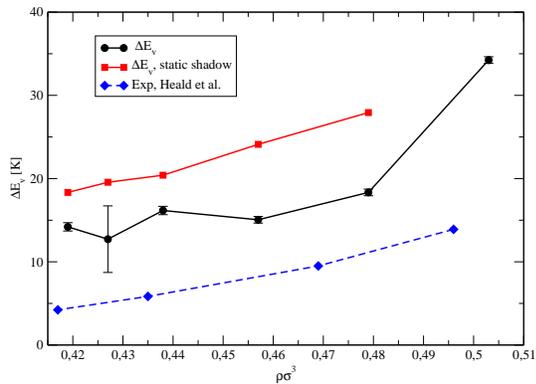}
\caption{(color online)
The vacancy formation energy $\Delta E_v$ as a function of the density using FSWF (black points) and 
by keeping the shadows fixed. Some experimental data from Ref. \cite{heald84} is also reported for comparison. 
See the text for details.}
\label{fig:enevac}
\end{center}
\end{figure}
The energies of the system with a vacancy are reported 
in table \ref{tab:ene}. The vacancy formation
energy obtained using Eq. \ref{eq:enevac} 
is given in Fig. \ref{fig:enevac}, where we included the same calculation using the static 
shadows (red points). In the latter case the shadow degrees of freedom are kept fixed on the
lattice sites so their effect is switched off. This corresponds to using a variational wavefunction 
of the antisymmetric Jastrow--Nosanow type. FSWF is more effective as is clear from the figure.
The vacancy formation energy computed by means of FSWF is larger than the experimental data (blue points) taken
from Ref. \cite{heald84} (see also Ref. \cite{grigorev07} and references therein).
The discrepancy can be attributed to several possible sources. First of all the calculation
might be affected by strong finite--size effects. In fact the effective concentration
of vacancies in the system is rather high ({\it{i.e.}} 1/N), and this might imply a contribution
to the vacancy--formation energy coming from a vacancy-vacancy interaction.
There is also additional room for improvement in the overall variational description.
For example, the parameters could be re-optimized in the presence of the vacancy, or
a more sophisticated version of the wavefunction including a local--density dependence
of the two--body correlations might be used\cite{pederiva94}. 

In conclusion, in this Letter we present a novel variational wavefunction to study fermionic systems with 
impurities.
We describe the Fermionic Shadow wavefunction that we  used to compute the equation of state of solid $^3$He 
in the b.c.c. phase, and the vacancy formation energy as a function of the density.
We stress the fact that using standard wavefunctions it is not possible 
to correctly study systems with the presence of defects or impurities like a vacancy.
In particular the theoretical study of $^3$He with vacancies  requires a correct description of
relaxation and tunneling effects that cannot be addressed by using standard forms of wavefunctions.
Within the variational framework we also computed the pair distribution functions between atoms and the density 
around a vacancy, and the calculation of other properties is possible, but this, as well as 
the technical improvement of our present method,  will be the subject of future work.
Using FSWF it is possible to study $^3$He with the presence of impurities of $^4$He as well
as the mixture of the two gases, and to move near the region where the solid and liquid phases start
to coexist. Work in these directions is in progress.

We thank G.V. Chester for useful discussions, C.J. Umrigar for providing us the Levemberg-Marquardt 
package used for the optimization of the wavefunction, and in particular we are indebted with F.
Operetto for help with the optimization procedure.
Calculations were partially performed on the BEN cluster at ECT* in Trento, under a grant for supercomputing 
projects, and partially on the HPC facility "WIGLAF" of the Department of Physics,
University of Trento.
This work was performed under the auspices of the U.S. Department
of Energy by Lawrence Livermore National Laboratory
under Contract DE-AC52-07NA27344.


\end{document}